\documentclass[aip,reprint,amsmath,amssymb]{revtex4-2}

\usepackage[utf8]{inputenc}
\usepackage[T1]{fontenc}
\usepackage{graphicx}
\usepackage{booktabs}
\usepackage{hyperref}
\usepackage{subcaption}

\begin{document}

\title{Tuning of superconducting and physical properties of
       Mo\textsubscript{1\textminus{}x}Si\textsubscript{x} thin films
       for Superconducting Nanowire Single-Photon Detector Applications}

\author{Oleh~V.~Zraichenko}
\email{zraichenko@ilt.kharkov.ua}
\affiliation{Department of Superconducting \& Mesoscopic Structures,
  B.~Verkin Institute for Low Temperature Physics and Engineering,
  NAS of Ukraine, Kharkiv, Ukraine}

\author{Oleksandr~O.~Leha}
\affiliation{Department of Superconducting \& Mesoscopic Structures,
  B.~Verkin Institute for Low Temperature Physics and Engineering,
  NAS of Ukraine, Kharkiv, Ukraine}

\author{Sergei V. Bengus}
\affiliation{Department of Superconducting \& Mesoscopic Structures,
  B.~Verkin Institute for Low Temperature Physics and Engineering,
  NAS of Ukraine, Kharkiv, Ukraine}

\author{Valerii~Yu.~Lyakhno}

\affiliation{Department of Superconducting \& Mesoscopic Structures,
  B.~Verkin Institute for Low Temperature Physics and Engineering,
  NAS of Ukraine, Kharkiv, Ukraine}

\date{}

\begin{abstract}
We investigate the normal-state electrical properties of ultrathin amorphous Mo$_{1-x}$Si$_x$ films deposited by DC magnetron co-sputtering over a broad range of compositions and thicknesses. Films with Si contents of $x = 0.14$–$0.5$ were deposited on Si/SiO$_2$ and Si / Si$_3$N$_4$ substrates (monocrystalline  Si with SiO$_2$ and S$_3$N$_4$ buffer layers respectfully); X-ray diffraction confirms an amorphous structure across the full composition range, while X-ray reflectometry is used to verify film thicknesses. The room-temperature sheet resistance ranges from 110 to 1990,$\Omega/\square$, corresponding to resistivities of 125–478,$\mu\Omega\cdot$cm. The sheet resistance generally increases with increasing Si content and decreasing film thickness, although the coupled variation of composition and thickness limits the quantitative separation of these effects. Additional measurements on thicker reference films show superconducting transition temperatures between 5.2 and 7.7,K, with $T_c$ decreasing monotonically at higher Si contents. A comparison of films on different substrate types shows no systematic difference between Si/SiO$_2$ and Si/Si$_3$N$_4$, while films on Al$_2$O$_3$ exhibit elevated sheet resistance at the lower thicknesses. These results characterize the accessible parameter space of our deposition process and provide a basis for the selection of film stoichiometry for detector fabrication.
\end{abstract}

\keywords{amorphous thin films, molybdenum silicide, sheet resistance,
          superconducting nanowire single-photon detectors,
          DC magnetron sputtering}

\maketitle

%% ---------------------------------------------------------------
\section{Introduction}
%% ---------------------------------------------------------------

Superconducting nanowire single-photon detectors
(SNSPDs)~\cite{Goltsman-2001} are among the best-performing
single-photon counters available today, combining high detection
efficiency, low dark-count rate, and low timing
jitter~\cite{Korzh-2020,Raj-2025}. The choice of superconducting
material is one of the key parameters for their optimization.
Besides the widely used NbN and NbTiN~\cite{Natarajan-2012},
NbRe nitride~\cite{Cirillo-et-al-2025}, Mo-Re and
MoC~\cite{Seleznev-2008,Trgala-2014}, amorphous silicides
such as WSi and Mo\textsubscript{1\textminus{}x}Si\textsubscript{x} have attracted attention: their lack of
long-range order makes the films compositionally homogeneous
over large areas, and their comparatively small superconducting
gap is favourable for the detection of low-energy
photons~\cite{Banerjee-2017,Zhang-2016}. WSi-based detectors have
demonstrated system detection efficiencies above
93\%~\cite{Marsili-2013}. Mo\textsubscript{1\textminus{}x}Si\textsubscript{x} was introduced as an SNSPD material
by Korneeva et al.~\cite{Korneeva-2014}; the single-photon response
of Mo\textsubscript{1\textminus{}x}Si\textsubscript{x} strips of different widths was subsequently studied
in~\cite{Korneeva-2020}. MoSi-based detectors have since reached
system detection efficiencies above 98\% at telecom
wavelengths~\cite{Reddy-2020}.

The electrical and superconducting properties of amorphous MoSi
depend strongly on both stoichiometry and
thickness~\cite{Banerjee-2017,Bosworth-2015}. For detector design,
the normal-state sheet resistance $R_\square$ is of particular
interest, as it sets the kinetic inductance of the nanowire and
thereby its electrical response~\cite{Annunziata-2010}. In the
ultrathin regime relevant for SNSPDs, small changes in thickness
or composition shift $R_\square$ considerably. Although the general
trends are known from the literature, every deposition system has
to be characterised individually before stoichiometry can be used
as a design parameter.

The aim of this work is therefore to establish the range of
normal-state resistance accessible with our co-sputtering process
over a broad composition window, $x = 0.14$--$0.5$, at
thicknesses of $2$--$50$\,nm. We begin with the structural
characterisation of the films, which forms the basis for the
transport analysis. We then examine the dependence of $R_\square$
and of the resistivity on composition and thickness, and finally
estimate the corresponding kinetic inductance and discuss the
implications for detector fabrication.

%% ---------------------------------------------------------------
\section{Experimental}
%% ---------------------------------------------------------------

\subsection{Film Deposition}

Amorphous $\mathrm{Mo}_{1-x}\mathrm{Si}_x$ thin films were
deposited by DC magnetron co-sputtering from separate Mo and Si
targets in an argon atmosphere at a working pressure of
$2\times10^{-3}$\,mbar. The base pressure prior to deposition was
$2\times10^{-6}$\,mbar. All depositions were performed without intentional substrate heating.

Film stoichiometry was controlled by independently adjusting the
deposition rates applied to each target: the Mo target was operated
at 1--4.5\,\AA/s and the Si target at 0.8--1.3\,\AA/s.
The Si fraction $x$ in the resulting films was varied in the range
$x = 0.14$--$0.5$. Films were deposited on Si/SiO$_2$
(230\,nm thermal oxide) and Si/Si$_3$N$_4$ substrates. A Si capping layer was deposited in situ immediately following deposition of each Mo\textsubscript{1\textminus{}x}Si\textsubscript{x} film to
prevent surface oxidation during storage and handling.

The Si fraction $x$ was calculated from the calibrated deposition
rates as  ~\cite{Erbe2024-bt}
\begin{equation}
  \frac{n_{mo}}{n_{si}} = \frac{d_{Mo}}{d_{Si}} \frac{\rho_{Mo}}{\rho_{Si}} 
  \cdot \frac{M_{Si}}{M_{Mo}},
  \label{eq:composition}
\end{equation}
where $r_{\mathrm{Mo}}$ and $r_{\mathrm{Si}}$ are the atomic
deposition rates proportional to the linear rates divided by the
respective atomic volumes. This yields $x = 0.14$--$0.5$.

\subsection{Thickness Calibration}

The deposition rates were calibrated using a quartz crystal
microbalance (QCM) prior to each deposition series. The film thickness
was calculated from the deposition time and the calibrated rate.
Selected samples were subsequently characterized by X-ray
reflectometry (XRR) to independently verify nominal thickness
values and to determine film density and surface roughness. XRR
data were fitted modelling a three-layer structure: substrate /
MoSi film / Si capping layer. The QCM-derived thicknesses were
confirmed within the experimental uncertainty. The film thicknesses in the
primary thin-film series ranged from 2 to 50 nm.

\subsection{Structural Characterisation}

The crystal structure of the deposited films was examined by X-ray
diffraction (XRD). No crystalline Bragg peaks were observed for
$\mathrm{Mo}_{1-x}\mathrm{Si}_x$ films across the full composition
range investigated ($x = 0.14$--$0.5$), confirming the amorphous
phase consistent with the literature~\cite{Banerjee-2017,Bosworth-2015}.
A crystalline Mo peak was observed only in pure Mo reference samples
($x = 0$), marking the onset of bcc crystallinity in the absence of
Si. Films with Mo content exceeding $\approx 80\%$ are known to adopt a bcc
structure with suppressed $T_c$~\cite{Kubo-1988}, which informed the
upper bound of the composition range investigated in this work.

\subsection{Resistance Measurements}

Room-temperature $R_\square$ was measured by the four-point probe method~\cite{Haldor-Topsoe-1968}. The width of the resistive transition $\delta T_c$ was used as a measure of film homogeneity; nonlocal four-probe effects in inhomogeneous superconducting films are discussed in~\cite{Zhitlukhina-2026}.
The correction factor for finite sample geometry was applied
according to the van der Pauw formalism, with values in the range
3.11--4.31 depending on sample dimensions. Measurements of the
critical temperature were performed on wire-bonded samples in a
four-wire configuration. Samples with well-defined composition and
thickness were included in the analysis~\cite{Zraichenko-2025}.

%% ---------------------------------------------------------------
\section{Results and Discussion}
%% ---------------------------------------------------------------

\subsection{Structure}

We start with the structural characterisation of the films.
No Bragg reflections are observed by XRD for any
$\mathrm{Mo}_{1-x}\mathrm{Si}_x$ film in the investigated
composition range, indicating an amorphous structure, in agreement
with previous reports~\cite{Banerjee-2017,Bosworth-2015}. A
crystalline Mo reflection appears only for the pure Mo reference
($x = 0$). For Mo-rich compositions above approximately 80\% Mo, a
bcc structure with suppressed $T_c$ has been reported for thicker
films~\cite{Kubo-1988}. The absence of such reflections in our most
Mo-rich films ($x = 0.143$) may be related to their small
thickness, which limits both the crystallisation and the diffracted
intensity; we therefore cannot exclude nanocrystalline inclusions at
this composition.

\subsection{Sheet Resistance and Resistivity}

The measured sheet resistance varies by more than an order of
magnitude across the sample set, reflecting the combined influence
of composition and thickness. To separate the geometric
contribution, we additionally consider the resistivity
\begin{equation}
  \rho = R_\square \cdot d.
\end{equation}

\subsection{Dependence on Composition}

\begin{figure}[htbp]
  \includegraphics[width=\columnwidth]{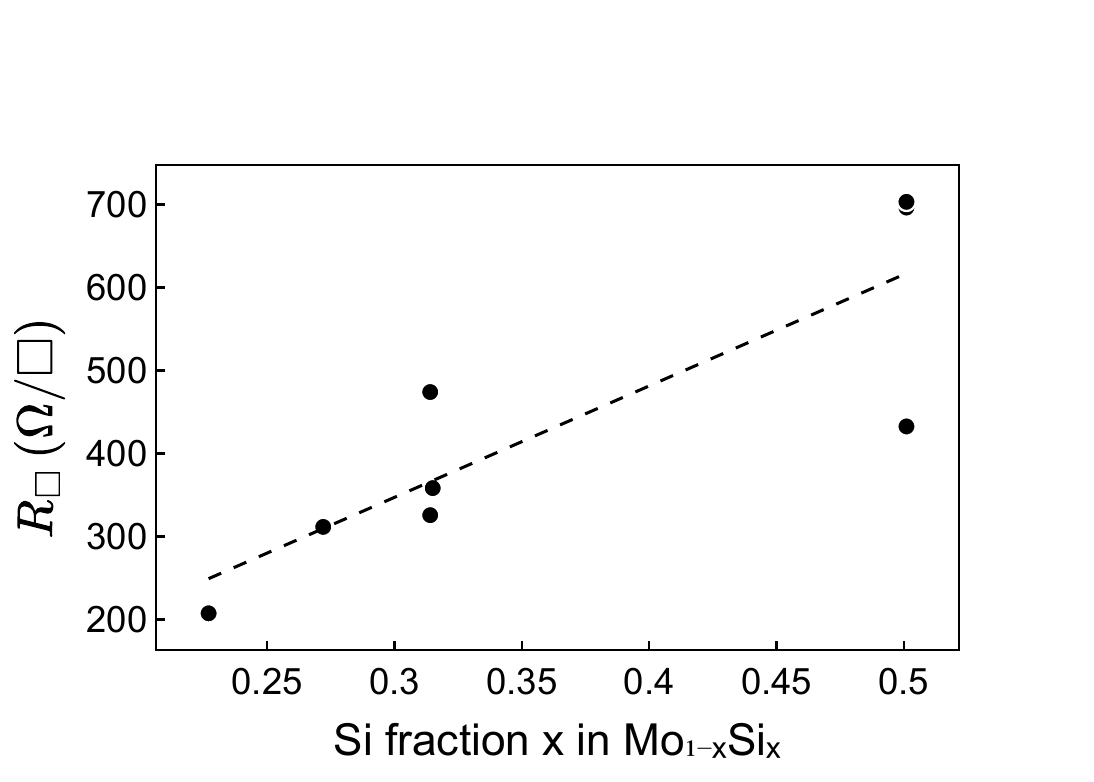}
  \caption{Sheet resistance $R_\square$ versus Si fraction $x$ for
    $\mathrm{Mo}_{1-x}\mathrm{Si}_x$ films with $d \approx 5$\,nm.
    Dashed line: exponential guide $R_\square = a\exp(bx)$.}
  \label{fig:Rs_vs_x}
\end{figure}

Figure~\ref{fig:Rs_vs_x} shows $R_\square$ as a function of Si
fraction for the thin films ($\approx 5$\,nm). $R_\square$ tends
to increase with Si content.  We attribute a large part of the
scatter around this curve to run-to-run variations in thickness and
deposition rate; a series with fixed thickness would be required to
extract the compositional dependence quantitatively.

The resistivity of all films ranges between 125 and
478\,$\mu\Omega\cdot\mathrm{cm}$. These values are in line with
those reported for amorphous MoSi~\cite{Banerjee-2017,Bosworth-2015}
and comparable to amorphous WSi films~\cite{Zhang-2016}, which
supports the interpretation of our films as strongly disordered
amorphous conductors~\cite{Mooij-1973}.

\begin{table}[htbp]
  \caption{Critical temperature $T_c$ of thicker reference films
    with nearly identical thickness, $d \approx 42$\,nm, as a
    function of Si fraction~$x$.}
  \label{tab:Tc_vs_x}
  \begin{ruledtabular}
    \begin{tabular}{cc}
      Si fraction & Critical temperature (K) \\
      \hline
      0.16 & 7.68 \\
      0.25 & 6.28 \\
      0.30 & 6.52 \\
      0.40 & 5.22 \\
    \end{tabular}
  \end{ruledtabular}
\end{table}

To assess the effect of composition on the superconducting
properties independently of thickness, we measured $T_c$ for a
series of thicker reference films with nearly identical thickness,
$d \approx 42$\,nm (Table~\ref{tab:Tc_vs_x}). Over the range
$x \approx 0.17$--$0.40$, $T_c$ decreases from about 7.7\,K to
about 5.2\,K. The two intermediate compositions, $x \approx 0.25$
and $0.30$, differ by about 0.2\,K, which is within the combined
uncertainty of the nominal composition and the transition width; we
therefore do not interpret this difference as a deviation from the
overall trend. The observed decrease of $T_c$ with Si content is
consistent with previous reports on amorphous MoSi. Banerjee et
al.~\cite{Banerjee-2017} found $T_c \approx 7.3$\,K for optimised
Mo$_{0.83}$Si$_{0.17}$ films of 20\,nm thickness, with $T_c$
falling to below 5\,K as the Si power was increased beyond the
optimum; the bulk extrapolation from the Finkel'stein model gives
$T_{\mathrm{co}} \approx 7.5$--$7.8$\,K for this composition.
Bosworth et al.~\cite{Bosworth-2015} similarly reported
$T_c \approx 7.4$\,K for Mo$_{0.83}$Si$_{0.17}$ thick films, and
a monotonic decrease with increasing Si content. Our values of
7.7\,K at $x \approx 0.17$ and 5.2\,K at $x \approx 0.40$ are in
good quantitative agreement with these results.

Because superconductivity in ultrathin films is further suppressed
with decreasing thickness, the values in Table~\ref{tab:Tc_vs_x}
represent an upper bound for $T_c$ of the 2--6\,nm films at the
same composition. Taken together with Fig.~\ref{fig:Rs_vs_x}, the
data show that increasing the Si content raises the normal-state
resistance and lowers $T_c$. For detector design, the composition
therefore trades a higher sheet resistance and kinetic inductance
against a lower operating temperature.

\begin{figure}
  \centering
  
  \includegraphics[width=0.49\columnwidth]{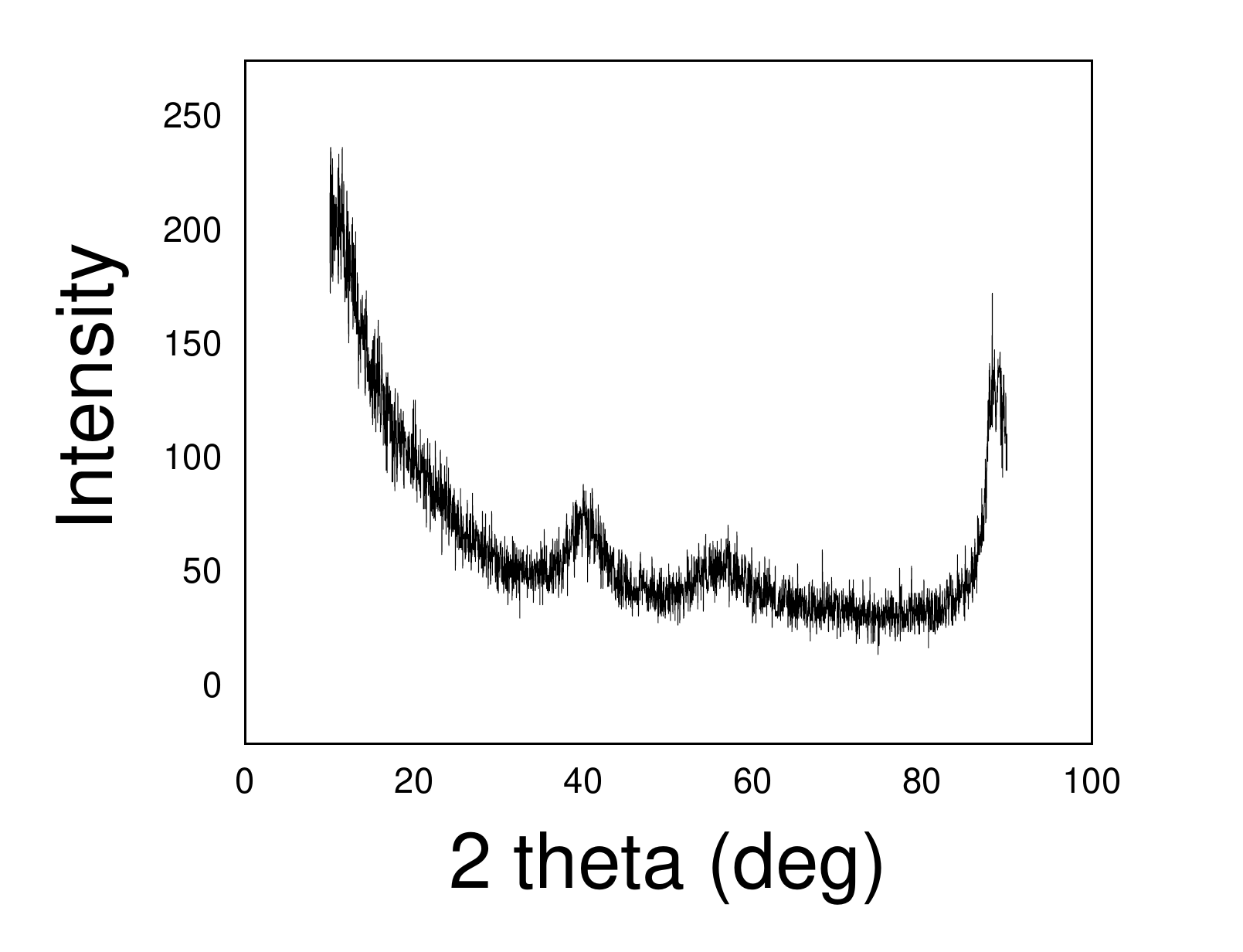}\hfill
  \includegraphics[width=0.49\columnwidth]{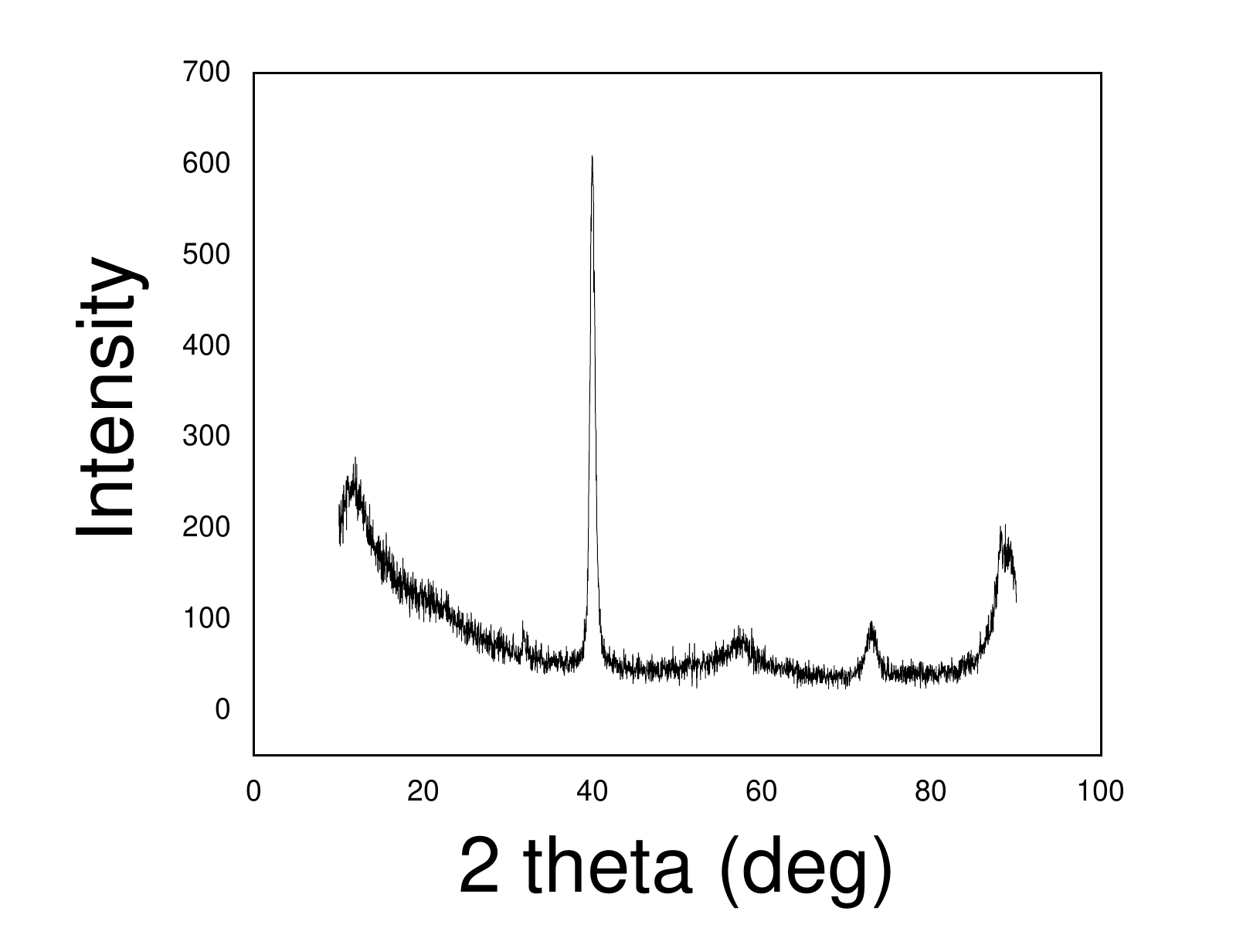}
  \caption{XRD patterns of (left) Mo$_{0.65}$Si$_{0.35}$ and (right) Mo films.}
  \label{fig:xrd_data}
\end{figure}

To verify that the XRD setup is sensitive to crystalline order, a pure Mo film (x = 0) was measured under the same conditions; it shows sharp bcc reflections at $2\theta\approx 40°$ and $\approx 73°$ ~\ref{fig:xrd_data}, whereas the $Mo_{0.65}Si_{0.35}$ film of comparable thickness to the reference series shows only a broad halo. The Tc values in ~\ref{tab:Tc_vs_x} therefore refer to a structurally different state from crystalline Mo.

\begin{figure}
  \centering
  \includegraphics[width=0.49\columnwidth]{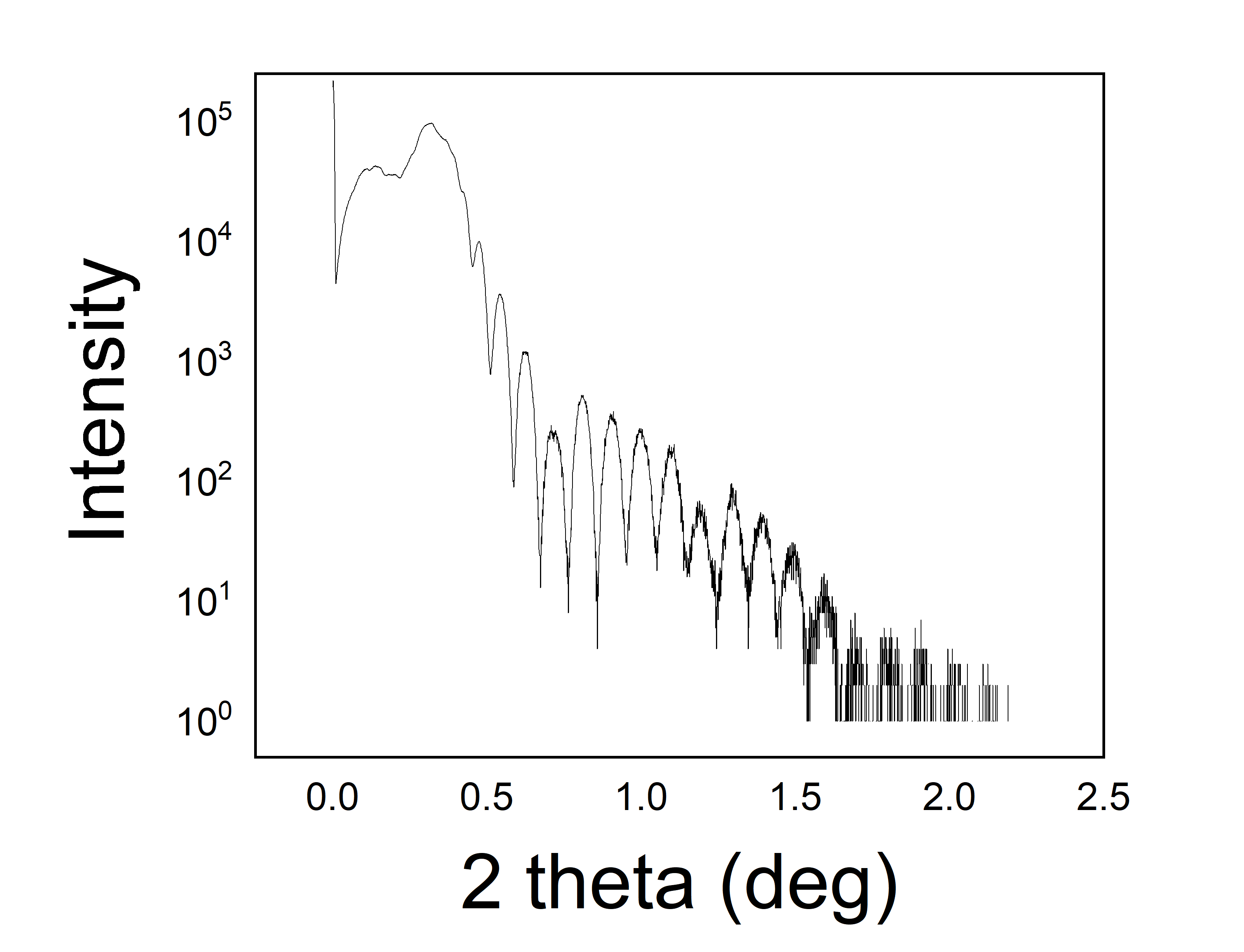}\hfill
  \includegraphics[width=0.49\columnwidth]{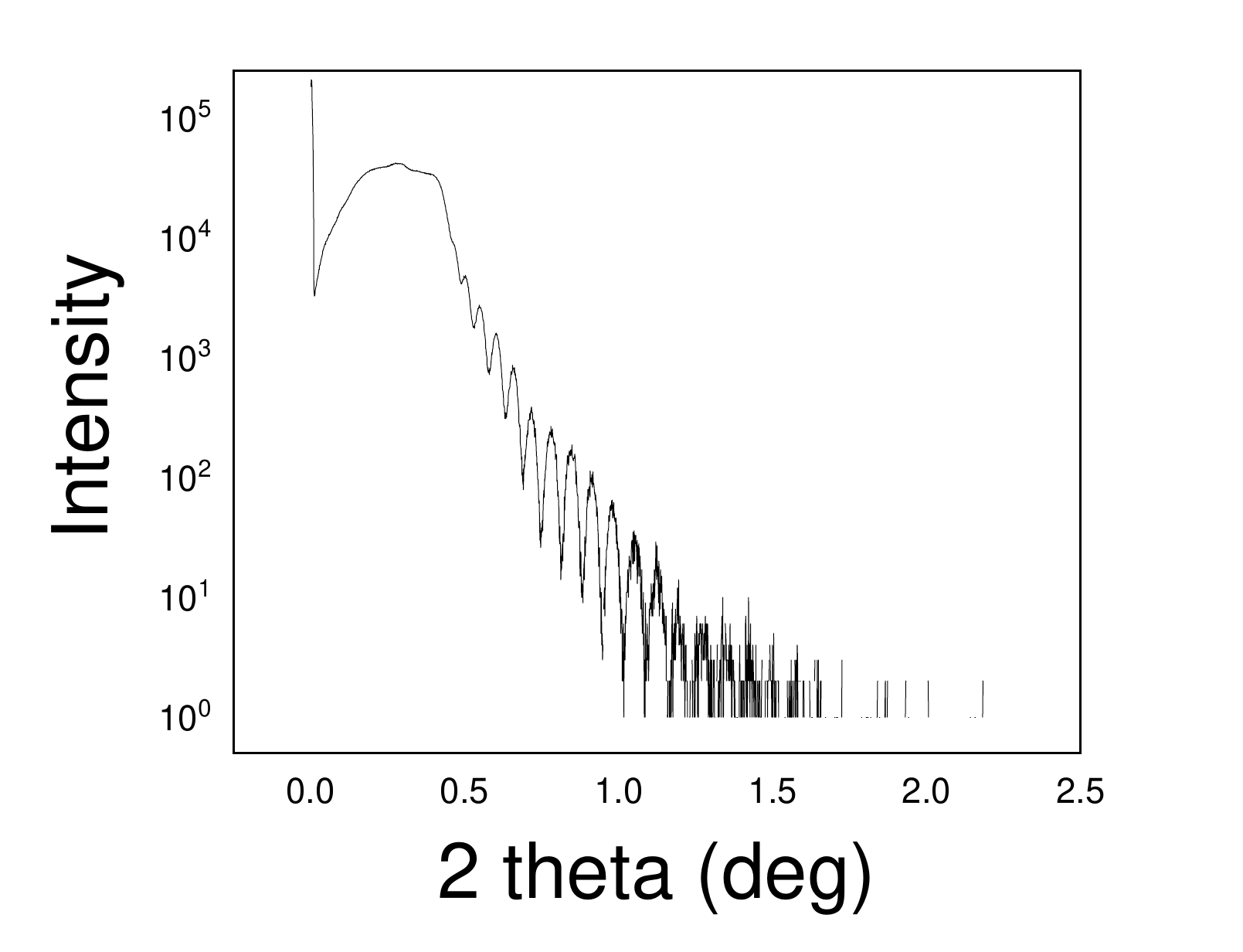}
  \caption{XRR patterns of (left) Mo$_{0.65}$Si$_{0.35}$ and (right) Mo films.}
  \label{fig:xrr_data}
\end{figure}

The curves were fitted using a three-layer model (substrate / Mo\textsubscript{1\textminus{}x}Si\textsubscript{x} / Si cap) with thickness, density and roughness as free parameters. Both the 41 nm $Mo_{0.65}Si_{0.35}$  film and the 72 nm Mo film show well-resolved Kiessig fringes ~\ref{fig:xrr_data}, indicating smooth interfaces.

These XRD and XRR data were presented as a poster at International Research and Practice Conference “Nanotechnology and Nanomaterials” (Nano 2026)~\cite{Zraichenko-2026}.

\subsection{Dependence on Thickness}

\begin{figure}[htbp]
  \includegraphics[width=\columnwidth]{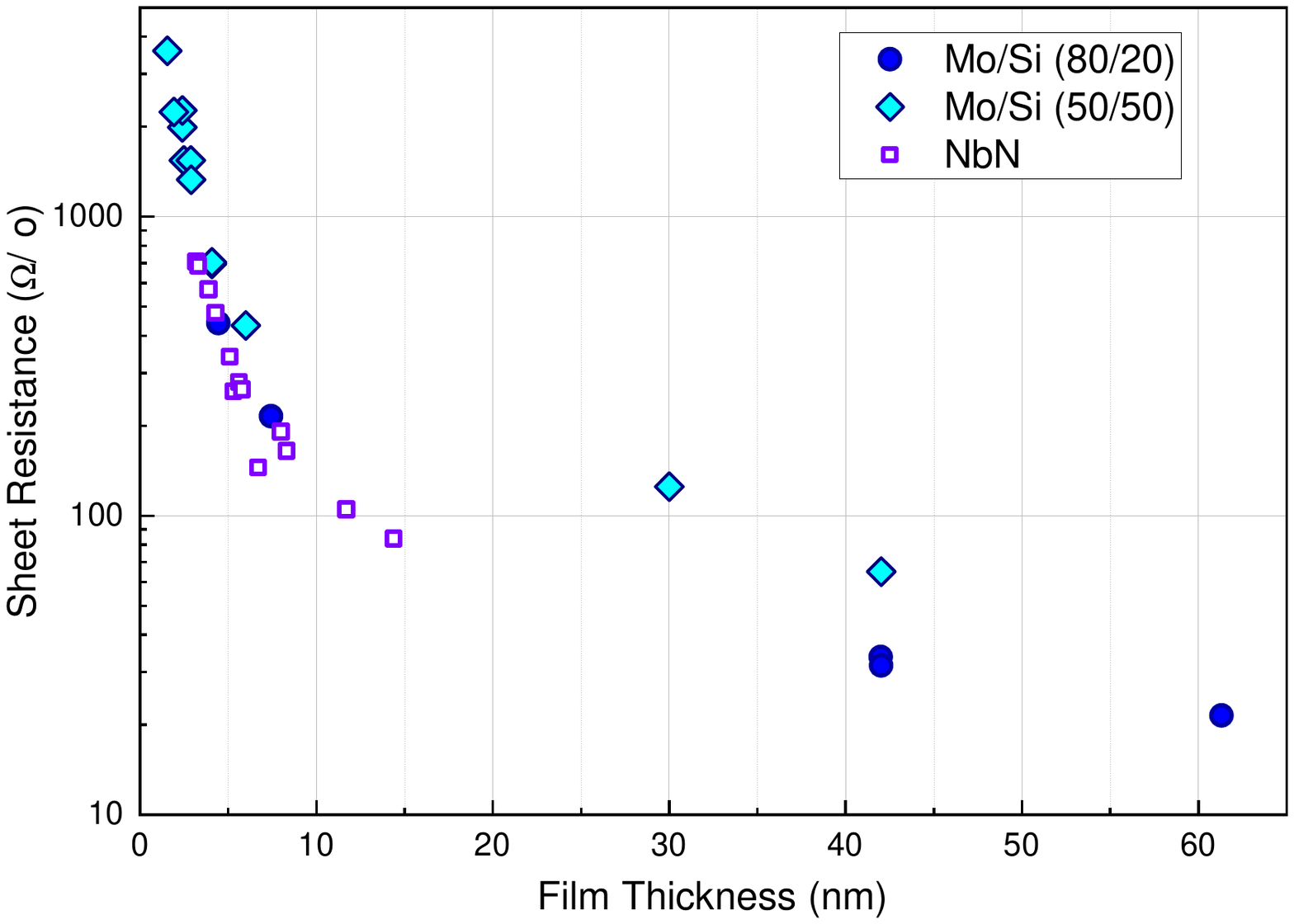}
  \caption{Sheet resistance $R_\square$ as a function of film thickness for
Mo$_{1-x}$Si$_x$ films with $x\approx0.2$ (Mo/Si 80/20, circles) and
$x\approx0.5$ (Mo/Si 50/50, diamonds), compared with NbN films (open squares).
The Mo$_{1-x}$Si$_x$ data are from this work (nominal QCM thicknesses);
the NbN data are taken from Ref.~\cite{Semenov-2009}. S (Color online)}
  \label{fig:Rs_summary}
\end{figure}

Figure~\ref{fig:Rs_summary} shows $R_\square$ as a function of film
thickness for $\mathrm{Mo}_{1-x}\mathrm{Si}_x$ films with $x\approx0.2$
and $x\approx0.5$, together with literature data for NbN
films~\cite{Semenov-2009}. For both compositions $R_\square$ increases
with decreasing thickness, from about $20$--$30\,\Omega/\square$ at
$d\approx 40$--$60$\,nm to several hundred $\Omega/\square$ at
$d\approx 5$\,nm and above $10^{3}\,\Omega/\square$ below about 3\,nm.
At a given thickness the $x\approx0.5$ films have a higher $R_\square$
than the $x\approx0.2$ films (by roughly a factor of two at
$d\approx 42$\,nm). In the range $d\approx 5$--$30$\,nm the increase is
close to $R_\square\propto 1/d$, i.e.\ to a thickness-independent
resistivity, whereas below $d\approx 3$\,nm it is considerably steeper
(Table~\ref{tab:substrate}). In the thickness range of $5$--$15$\,nm
the $R_\square$ values of our films are of the same order as those of
NbN of comparable thickness.

Within this series, $\rho$ ranges from 273 to
478\,$\mu\Omega\cdot\mathrm{cm}$, i.e.\ it varies by a factor of
about 1.7, which is small compared with the change in $R_\square$.
The data are thus broadly compatible with an approximately inverse
dependence of $R_\square$ on $d$. The residual variation of $\rho$
could originate from thickness-dependent disorder, surface and
interface contributions, or from small deviations in stoichiometry,
which we cannot separate with the present data. A dedicated
thickness series at independently verified composition is needed to
resolve this.

\subsection{Dependence on Substrate}

To examine whether the substrate affects the normal-state sheet resistance, Table~\ref{tab:substrate} compares $R_\square$ values for $\mathrm{Mo}_{1-x}\mathrm{Si}_x$ films with $x \approx 0.50$ deposited on three substrate types: Al$_2$O$_3$, Si/SiO$_2$, and Si/Si$_3$N$_4$. The nominal film thickness ranges from 1 to 30 nm. The sheet resistance values for films deposited on Si/SiO$_2$ and Si/Si$_3$N$_4$ are relatively close across the investigated thickness range. Films on Al$_2$O$_3$ exhibit higher sheet resistance at intermediate thicknesses (approximately 1.54–2.9 nm), whereas at 1 nm the sheet resistance is lower than that measured on the other substrates. At thicknesses of 5 and 30 nm, the differences between the three substrate types are comparatively small. These results suggest that substrate-related differences may become more pronounced in the ultrathin regime, although the present dataset does not establish a systematic dependence across the entire thickness range.

Over most of the thickness range the data for all three substrates
follow the $R_\square \propto d^{-1}$ guide, consistent with a
bulk-like, thickness-independent resistivity. Films on Si/SiO$_2$
and Si/Si$_3$N$_4$ agree with each other within the experimental
scatter across the entire range investigated, and no systematic
difference between these two substrate types is observed. This is
consistent with the expectation that the normal-state transport in
amorphous Mo\textsubscript{1\textminus{}x}Si\textsubscript{x} is governed by the bulk film rather than by the
substrate or the film--substrate interface.

Films on Al$_2$O$_3$, by contrast, show a systematically
elevated $R_\square$ at thicknesses below approximately 3\,nm,
while converging with the SiO$_2$ and Si$_3$N$_4$ data for
$d \gtrsim 6$\,nm. This behaviour may be attributed to enhanced
interface scattering at the MoAl$_2$O$_3$ interface, or to
incomplete film coalescence in the thinnest films on this substrate;
the present data do not allow these two mechanisms to be separated.
Since the films used for the transport analysis in this work all
have $d \geq 2.4$\,nm and are deposited on SiO$_2$ or
Si$_3$N$_4$, the Al$_2$O$_3$ interface effect does not affect the
conclusions drawn above.

\begin{table}
\caption{Sheet resistance $R_\square$ (in $\Omega$) of Mo$_{1-x}$Si$_x$ films ($x\approx0.50$) of different nominal thicknesses on three substrate types. The NbN data are taken from Ref.~\cite{Semenov-2009}.}
\label{tab:substrate}
\begin{ruledtabular}
\begin{tabular}{lccc}
 & \multicolumn{3}{c}{Substrate} \\
\cline{2-4}
$d$ (nm) & Al$_2$O$_3$ & Si/SiO$_2$ & Si/Si$_3$N$_4$ \\
\hline
30.0 & 136   & 125   & 133   \\
5.0  & 908   & 855   & 864   \\
2.9  & 1766  & 1329  & 1202  \\
2.4  & 2744  & 2269  & 2281  \\
1.92 & 2360  & 2241  & 2070  \\
1.54 & 4638  & 3589  & 3868  \\
1.0  & 16350 & 18518 & 17777 \\
\end{tabular}
\end{ruledtabular}
\end{table}

\subsection{Implications for SNSPD Fabrication}

The practical relevance of $R_\square$ for detector design lies in
its relation to the kinetic inductance~\cite{Semenov-2005}. 
In the dirty limit and at $T \ll T_c$~\cite{Tinkham-2004}, 
the kinetic inductance per square is~\cite{Annunziata-2010}
\begin{equation}
  L_K/\square = \frac{\hbar R_\square}{\pi \Delta},
  \label{eq:Lk}
\end{equation}
where $\Delta$ is the superconducting energy gap. For the films
with $x = 0.27$--$0.32$ and $d \approx 5$--$6$\,nm, the measured
sheet resistance is approximately 200--360\,$\Omega/\square$.
Using $T_c \approx 6.3$--$6.5$\,K measured for thick films with
$x \approx 0.25$--$0.30$ (Table~\ref{tab:Tc_vs_x}) and the BCS
relation $\Delta = 1.76\,k_B T_c \approx 0.95$--$0.99$\,meV,
Eq.~(\ref{eq:Lk}) yields $L_K \approx 42$--$79$\,pH/$\square$.
Since $T_c$ of the ultrathin films is lower than that of the thick
reference films, these values represent a lower bound for the
kinetic inductance.

%% ---------------------------------------------------------------
\section{Summary and Conclusion}
%% ---------------------------------------------------------------

In summary, we characterised the normal-state resistance of 21
ultrathin amorphous $\mathrm{Mo}_{1-x}\mathrm{Si}_x$ films
deposited by DC magnetron co-sputtering with $x = 0.14$--$0.5$
and $d = 2$--$6.1$\,nm. XRD indicates an amo rphous structure
over the investigated composition range, and XRR confirms the
QCM-derived thicknesses. The sheet resistance spans roughly
110--1990 $\Omega/\square$ and increases with Si content and
decreasing thickness, whereas the resistivity remains within
125--478 $\mu\Omega\cdot\mathrm{cm}$, consistent with literature
values for amorphous MoSi. For the $x \approx 0.50$ series, the
thickness dependence is broadly compatible with
$R_\square \propto 1/d$. The roughly estimated kinetic inductance of
approximately 42--79\,pH/$\square$ for films with $x = 0.27$--$0.32$
lies in a range relevant for SNSPDs.

%% ---------------------------------------------------------------
\begin{acknowledgments}
The authors gratefully acknowledge the support of the National
Research Foundation of Ukraine (NRFU) under grant No.~22.07/0044.
This work was supported in part by IEEE ``Magnetism for Ukraine
2025--2026'', project number~9918.
\end{acknowledgments}

\bibliography{refs}

\end{document}